\documentstyle[12pt,amssymb,amscd]{amsart}  

\theoremstyle{remark}

\theoremstyle{definition}

\numberwithin{equation}{subsection}

\def\bea{\begin{eqnarray}}
\def\eea{\end{eqnarray}}
\def\beq{\begin{equation}}          
\def\eeq{\end{equation}}            
\def\nn{\nonumber}                  

\newcommand{\bbC}{{\Bbb C}}

\newcommand{\bbR}{{\Bbb R}}

\newcommand{\bbZ}{{\Bbb Z}}
\newcommand{\cF}{{\cal F}}

\newcommand{\cO}{{\cal O}}

\newcommand{\cL}{{\cal L}}

\newcommand{\cD}{{\cal D}}

\newcommand{\cA}{{\cal A}}
\newcommand{\cB}{{\cal B}}

\newcommand{\cT}{{\cal T}}
\newcommand{\cQ}{{\cal Q}}

\newcommand{\shHom}{\underline{\operatorname{Hom}}}

\newcommand{\pr}{\operatorname{pr}}
\newcommand{\ad}{\operatorname{ad}}

\catcode`\@=11
\def\marginnote#1{}

\newcount\hour
\newcount\minute
\newtoks\amorpm
\hour=\time\divide\hour by60
\minute=\time{\multiply\hour by60 \global\advance\minute by-\hour}
\edef\standardtime{{\ifnum\hour<12 \global\amorpm={am}%
        \else\global\amorpm={pm}\advance\hour by-12 \fi
        \ifnum\hour=0 \hour=12 \fi
        \number\hour:\ifnum\minute<10 0\fi\number\minute\the\amorpm}}
\edef\militarytime{\number\hour:\ifnum\minute<10 0\fi\number\minute}

\def\draftlabel#1{{\@bsphack\if@filesw {\let\thepage\relax
   \xdef\@gtempa{\write\@auxout{\string
      \newlabel{#1}{{\@currentlabel}{\thepage}}}}}\@gtempa
   \if@nobreak \ifvmode\nobreak\fi\fi\fi\@esphack}
        \gdef\@eqnlabel{#1}}
\def\@eqnlabel{}
\def\@vacuum{}
\def\draftmarginnote#1{\marginpar{\raggedright\scriptsize\tt#1}}

\def\draft{\oddsidemargin -.5truein
        \def\@oddfoot{\sl preliminary draft \hfil
        \rm\thepage\hfil\sl\today\quad\militarytime}
        \let\@evenfoot\@oddfoot \overfullrule 3pt
        \let\label=\draftlabel
        \let\marginnote=\draftmarginnote
   \def\@eqnnum{(\theequation)\rlap{\kern\marginparsep\tt\@eqnlabel}%
\global\let\@eqnlabel\@vacuum}  }

\newcounter{app}
\newcounter{sapp}[app]

\newcommand{\app}[1]{
\refstepcounter{app}{\noindent\Large\bf Appendixes
 \ #1 \par \vspace{5mm}}
\setcounter{equation}{0}
\def\theequation{\Alph{app}.\arabic{equation}}}
\def\thesapp{\Alph{app}.\arabic{sapp}}
\newcommand{\sapp}[1]{\par \refstepcounter{sapp}{\noindent\large\bf \thesapp
\ #1 \par \vspace{3mm}}
\def\theequation{\Alph{app}.\arabic{equation}}}

\begin{document}

\begin{titlepage}

\setlength{\baselineskip}{2\baselineskip}
\vspace{0.3in}
\begin{flushright}
 ITEP-TH-29/01\\
\end{flushright}
\vspace{10mm}

\title{Courant Algebroids}

\author{Paul Bressler}
\author{Alexander Chervov} 

\address{}
\email{chervov@gate.itep.ru}
\email{bressler@math.arizona.edu}


\maketitle
\begin{abstract}

This paper  is devoted to studying some properties of the Courant
algebroids: we  explain the so-called
"conducting bundle construction" and
use it to attach the Courant algebroid to Dixmier-Douady gerbe
(following ideas of P. Severa).
We remark that WZNW-Poisson condition of
Klimcik and Strobl  (math.SG/0104189)
is the same  as Dirac structure in some
particular Courant algebroid.
We propose the construction of the Lie
algebroid on the loop space starting
from the Lie algebroid on the manifold
and conjecture that this construction applied
to the Dirac structure above should give
the Lie algebroid of symmetries in the WZNW-Poisson
$\sigma$-model, we show that it is indeed true
in the particular case of Poisson $\sigma$-model.

\end{abstract}

\end{titlepage}

\tableofcontents

\section{Introduction}

In this note we study some properties of Courant algebroids
and argue their natural connection to gerbes, WZNW-Poisson
manifolds and possibly Poisson $\sigma$-model.

Courant algebroids introduced in \cite{Cour}
are certain kinds of algebroids  for which
the conditions  antisymmetricity, Leibniz rule and the Jacobi identity
for the bracket
are relaxed in a certain way.
Original motivation of \cite{Cour} was to characterize
Poisson and symplectic structure in the same manner,
such that the graph of the map given by the Poisson structure
 from the 1-forms ( from the vector fields in the symplectic case)
to the direct sum of 1-forms and vector fields
is subalgebra with respect to the Courant bracket.
Further investigation \cite{LWX} showed that
the Courant algebroid is very natural setup
for the construction of bialgebroids.
Also Courant algebroids are relevant to studying
dynamical $r$-matrices \cite{LX}
and have various other applications
(see \cite{Roy} and references their in).

Let us recall the Courant bracket
on $T^* M \oplus T M$:

\bea
[\alpha_1 + \xi_1, \alpha_2 + \xi_2]=
(\cL_{\xi_1} \alpha_2 - \cL_{\xi_2} \alpha_1+
\frac{1}{2}d(\iota_{\xi_2 }\alpha_1 -\iota_{\xi_1 }\alpha_2))
+[\xi_1,\xi_2] \label{Cour-br1}
\eea

The bracket is antisymmetric, but the Leibniz
rule and Jacobi identity are not true,
but the anomalies for them are easily expressible in terms
of the pairing:

\bea
<\alpha_1 + \xi_1, \alpha_2 + \xi_2>=
(\iota_{\xi_2 }\alpha_1 +\iota_{\xi_1 }\alpha_2)
\label{pair}
\eea

In our text (except for the section \ref{s-WZ-P}) we will use slightly
modified bracket:

\bea
[\alpha_1 + \xi_1, \alpha_2 + \xi_2]=
(\cL_{\xi_1} \alpha_2 - \cL_{\xi_2} \alpha_1+
d(\iota_{\xi_2 }\alpha_1))
+[\xi_1,\xi_2]=  \nn \\
= \cL_{\xi_1} \alpha_2 - \iota_{\xi_2} d\alpha_1+
[\xi_1,\xi_2]
\label{Cour-br2}
\eea

This bracket is not antisymmetric but one has the left Jacobi identity and
right Leibniz rule for it. This was discovered by Y. Kosmann-Schwarzbach, P.
Severa and one of the authors (P.B.) independently (unpublished).
 The
brackets are connected to each other by adding the $\frac{1}{2}<,>$.

{\it Definition:} let us call the bundle
$T^*M \oplus TM$ with the bracket
\ref{Cour-br2} and anchor map $\pi : \alpha + \xi \mapsto \xi$
the {\it standard Courant} algebroid.

We refer to \cite{LWX} or to the section \ref{Cour-def} of the present
paper for the general definition
of Courant algebroids.

The paper organized as follows in
section \ref{prelim} we recall some basic definitions.

In section \ref{Cond} we realize in details
the "conducting bundle"  construction
(idea of which is due to P. Severa).
The construction is the following:
to any extension

\bea
0 @>>> \cO @>>> \widetilde\cT @>{\sigma}>> \cT @>>> 0
\label{ext}
\eea
we  associate the Courant algebroid
in a canonical and functorial way.
The main property of this construction
is the following (see section \ref{spl}):
assume there was an algebroid structure on
the extension above
then one has canonical trivialization
of the obtained Courant algebroid.
In smooth situation (opposite to holomorphic) all extension
can be endowed with the bracket,
hence all Courant algebroids obtained this way are the standard
ones.
But even in  smooth situation the
construction is not useless.
Following P.Severa, this construction can be used to
 attach a Courant algebroid to
a gerbe. Moreover morally speaking the relation of gerbes
to Courant algebroids is the same as of line bundles to their Atiyah
algebroids. But the precise formulation of this analogy is yet unknown.


In section \ref{exCour}
we study the properties of the  exact Courant algebroids,
this is subclass of Courant algebroids
which are the extensions
of the type:

\bea
0 @>>> \Omega^1 @>>> \cQ @>{\pi_\cQ}>> \cT @>>> 0
\eea

This kinds of algebroids can be obtained by the
construction of the  previous section.
P. Severa
noticed the following:
deformations of such algebroids
are governed by  the closed 3-forms.
This is quite striking fact.
He also proposed to mimic the differential geometry
for the Courant algebroids
i.e. to introduce the connections,
their curvature and prove the
basic properties analogous to the classical
situation e.g. the connections
are the affine space over the $2$-forms,
the curvature is closed 3-form and etc.
In particular using these properties
one can see  that there
is natural structure of $\cO$-module
on the Courant algebroid obtained
by the previous construction.


At the end of this section we make
 a remark that WZNW-Poisson condition
of Klimcik and Strobl \cite{Klim}
coincides with Dirac structure in the
Courant algebroid twisted by the 3-form
(this was discovered independently by  P.Severa and A.Weinstein \cite{SW}).

 Section \ref{LoopCour} is motivated by the
papers \cite{Klim,LO} and we try to give
more geometric understanding of some of
their constructions.
Consider  the
diagram:
\bea
\begin{CD}
 Map(X,Y) & \times & X  @>ev>> Y \\
 @V{\pr}VV     \\
Map(X,Y) &&       \\
\end{CD}
\eea

We will show that if $A$ is an algebroid on $Y$
then there $pr_* ev^* A$ has a natural structure of algebroid,
considering the $T^*M$ as an algebroid with
 the bracket constructed from  Poisson bracket on $M$
and considering its pullback $pr_* ev^*T^*M$ on the loop space $LM$ we see
that this algebroid is  the same as algebroid considered in \cite{LO} which
plays the role of symmetries in Poisson $\sigma$-model. Analogous algebroids
were considered in \cite{CFel,CF}. We also hope  that for a Courant
algebroid on $M$ one can naturally construct Lie algebroid on $LM$ this
construction should be agreed with the construction of line bundle on $LM$
from gerbe on $M$ and   the transgression of
characteristic class of gerbe to characteristic class of line bundle on
$LM$.



{\bf Remark on Notations.}\\
In this work will denote by $\cO$ functions on our manifolds.
Most of the considerations works for analytic or algebraic  as well as for smooth
functions. The basic situation in our paper that $\cO$ are smooth functions.
We are sorry, that it may cause some inconvenience for those readers,
who get used to think about $\cO$ only as about analytic or algebraic
functions.
The same words should be said about our notation $\cT$ - it is tangent bundle,
basic example is smooth tangent bundle for smooth manifold, but most
considerations works well for the holomorphic tangent bundle as well.

{\bf Remark on Further Developments.}\\
This paper was basically finished in summer 2001.
Since that time there have been some further developments in the subject.
One of the authors (P.B.) in  \cite{VA1} proposed the analogy
between Courant algebroids and  ``vertex algebroids'' i.e.
algebroids that appeared in \cite{Sch1, Sch2} and which
 gives rise to the chiral de Rham complex
 and also he gave a ``coordinate free'' construction
and proved the uniqueness of the vertex algebroid.
The paper \cite{SW} contains many important ideas related to the subject of
present work.

{\bf Acknowledgements.}\\
The first part of this work owes its existence to P. Severa,
who expressed most of the ideas presented there in his unpublished letters to A. Weinstein.
We hope that this paper  clarifies and makes accessible a part of his ideas.
This work was done during the
authors stay in the University d'Angers to which we are grateful
for hospitality and excellent working conditions.
We are grateful to V. Rubtsov, N. Markarian  for useful discussions.
The second author is thankful to M. Crainic, S. Gukov, D. Osipov
for useful correspondence and discussions, to A. Gerasimov for sharing
his unpublished ideas and especially indebted to A. Kotov
for stimulating and suggestive discussions.
The work of A.C. was partially supported by the grant
RFBR  01-01-00546 and
Russian President's grant 00-15-99296


\section{Preliminary definitions} \label{prelim}

For the convenience of the reader let us recall
definitions of Lie-algebroids, Atiyah algebroid, gerbes,
Courant algebroids.

\subsection{Lie algebroids}

{\it Definition:}
Let $M$ be a smooth (complex) manifold.
A {\it Lie algebroid}
 is a vector bundle $A$ endowed with a Lie algebra
bracket $[,]_{A}$ on the space of sections $\Gamma (A)$ and
an anchor map $p: A\to TM$, which induces a Lie algebra homomorphism
$\Gamma (A)\to Vect(M)$ satisfying the Leibniz rule
\[
[X, fY]_{A}=f[X, fY]_{A}+(a(X)f)Y
\]
We refer to \cite{Mac} for the comprehensive study of Lie
algebroids and groupoids, see also \cite{Weins-book}.
In a sense Lie algebroid is a generalization
of the notion of tangent bundle, which is
the  example of Lie algebroid which
should be kept in mind first of all.
The natural examples of Lie algebroids are
Atiyah algebroid and the algebroid structure on $T^*M$
constructed from Poisson bracket on $M$.
The notion of de-Rham complex, Lie derivatives
and etc. holds identically  true for any
Lie algebroid. One is refered to
\cite{ELW} for the interesting studies
of this theory.

\subsection{Atiyah algebroid}

Atiyah algebroid is the natural example of Lie-algebroid
constructed from any bundle on $M$
(or speaking algebraically
from projective module over some commutative  algebra $A$.)
We will use this construction in order to explain the
relation of Courant algebroid to gerbes.
Informally speaking, Courant algebroid should be an analogue
or somehow related
to Atiyah algebroid,  not for the bundle,
but for the gerbe, but the precise
formulation is not yet known.

The idea of construction of Atiyah algebroid goes back to the paper
\cite{At}. It was studied for example in \cite{BS,Ru}.
For very nice modern exposition of Atiyah classes
see first paragraphes of \cite{Ka}.
The construction associates to any vector bundle a
Lie algebroid, which morally plays the role
of the Lie algebra of group of local automorphisms
of the bundle. The construction is functorial with respect to automorphisms of vector bundle.
The Atiyah algebroid is the space of
derivations of module of sections of bundle.
Let us give formal definitions below.

{\it Definition:} let us call the {\it derivation
of the module $P$} over some commutative ring $A$
the pair of maps $d: A \to A$ and $\bar d :  P \to P$
such that $d$ is derivation of algebra $A$
and $ \bar d$ is derivation of $P$, which lifts $d$ i.e.
$\bar d ( a p)= a \bar d p + d(a) p$, for $a\in A$ and $p \in P$.

Let us consider some manifold $M$ and the vector bundle $F$
over it. Let us denote the algebra of functions
on $M$ by $A$ and the module of sections of $F$ by $P$.

{\it Definition:} {\it Atiyah algebroid $A(F)$ (also denoted $Der(F)$} of the
bundle $F$  is
the set of all derivations of the module $P$.

(In algebraic or holomorphic situation
we should add the word sheaf everywhere).

{\it Lemma:}Obviously the following is true:
\begin{enumerate}
\item
$A(F)$ is naturally $A$-module.
\item  $A(F)$ is naturally Lie algebra,
with respect to commutator of maps.
\item the "anchor" map $\bar d \to d $
is Lie algebra homomorphism
\item
There is the following
exact sequence of Lie algebras.

\bea
End(P) @>>> A (F) @>>> \cT_M \label{der-seq}
\eea

where $End(P)$ is the space $A$-linear endomorphisms of $P$,
and $\cT_M$ is the module of vector fields (derivations) on M.

\end{enumerate}

{\it Warning:} note the derivations of $A$ considered
as an algebra and as a module over itself
are not the same, there is
an exact sequence

$ A @>>> Der_{module} A @>>> Der_{algebra} A$


{\it Remark:} the {\it connection} on the bundle
$F$ has a nice interpretation as  a splitting
$\nabla: \cT_M @>>> A (F)$
of the  sequence \ref{der-seq}. We mean the splitting
is homomorphism of $A$-modules. The curvature
is obviously the measure how far is  the splitting
from the homomorphism of Lie algebras.
The Bianchi identity   follows
from Jacobi  identity.
(See \cite{Al} for the  nice expositions of this matters).

{\it Lemma:} the association
$F \to A(F)$ is
functorial with respect to automorphisms of $F$.

Actually, if $\Phi$ is an automorphism of $A$-module $M$
and $d$ is a derivation of $M$, then $\Phi d \Phi^{-1}$ is again
derivation of $M$.

{\it Example:}
let us consider the linear bundle $L$ over $M$,
then its Atiyah algebroid is algebroid of the type:
$\cO @> i >> A(L) @> p >> \cT_M$

One knows that such algebroids are classified
by the $H^2(M)$.

{\it Example:} If an invertible function $g$ gives an automorphism of $L$,
then induced automorphism of $A(L)$ is given by the 1-form $g^{-1}dg=d log(g)$.

{\it Lemma:} The association of $L @>>> A(L)$
on the level of cocycles corresponds to the
map $f \mapsto d log (f)$, where $f \in H^1(\cO ^*)$
(i.e. in Cech description it is invertible function
on the intersection of charts)
and $d log (f)$ is closed 1-form on the intersection
of charts i.e. it corresponds to Cech-deRham description
of $H^2(M)$.

{\it Remark:}
Let us make a remark that
the extension $\cO @>>> A(L) @>>> \cT$
is not trivial extension of coherent
sheaves in general (see \cite{At}) (we mean holomorphic situation,
in smooth case all extensions are trivial).
The triviality of extensions means that
there exists holomorphic connection
(because as it was explained above
the splitting of this extension is
the same as connection on $L$.)

\subsection{Lie Algebroid on $T^*M$ from Poisson bracket on $M$.}
\label{Pois-alg}

Let $M$ be a manifold and $\pi$ be Poisson bracket on
it.
Then there is natural structure of the Lie algebroid
on the $T^*M$.
We will use it in  relation to WZNW-Poisson $\sigma$-model.

The Poisson bracket on functions
can be naturally extended
to the bracket on 1-forms
 \cite{Pois} defined by the rules $ [df, dg]=d \{ f,g \} $,
and Leibniz rule for functions$ [hdf, dg]=\{ h,g \} df+h[df,dg]$;
$[f, dg]= \{ f,g \} $.
The bracket can be written
 explicitly as follows:
$[ \alpha, \beta] =  \cL_{<\pi,\alpha>}\beta
-  \cL_{<\pi,\beta>}\alpha  -
 d < \pi, \alpha  \wedge \beta> $.


This bracket supplied with the anchor
map $p : T^*M \to TM$, by the rule $\alpha \mapsto
<\pi, \alpha>$ gives the $T^*M$ the structure of
Lie algebroid.

Where we denoted by $<\pi, \alpha>$ a vector which is the contraction
of bivector $\pi$  and 1-form $\alpha$.
(One can check that the anchor is really a
homomorphism of Lie algebras).

\subsection{Gerbes}
Gerbes were introduced by Giraud \cite{Gi} in early seventies.
Recently they were found to be useful in different
problems of mathematics and physics (see \cite{Br1, Br2, Hit,Gerbes}).
Informally speaking gerbes (more precisely Dixmier-Douady gerbes)
 are higher analogues of
the linear bundles, i.e. line bundles are classified
ny the $H^2(X,\bbZ)$ and gerbes are the objects classified
by $H^3(X,\bbZ)$.
There two ways to treat gerbes one is
more concrete and is similar to defining
the line bundles by the gluing functions on each chart
(see \cite{Hit}). Another is more abstract - the
language of sheaves of categories (see \cite {Br1}).
We refer the reader to section 1 of \cite{Br2}
for short and concise discussion of the material
and for the comparison of the two approaches.

{\it Definition:} a sheaf of groupoids
is called {\it Dixmier-Douady gerbe} (or DD-gerbe)
if it satisfies the following:

1) all objects $\cF (U)$ are locally isomorphic;

2) for any $x\in M$ there is an open set $U $ such
that $\cF (U)$ is not empty;

3) for any object $P$ of  $\cF (U)$, the
automorphisms of $P$ are exactly the smooth
functions $U \to \bbC^*$.

{\it Proposition \cite{Br1}:} The DD-gerbes are classified by the
$H^{3}(M,\mathbb{Z})$.

In analogy with the theory of line bundles
there are the notions of connection on the gerbe,
but there are two kinds of connections:
$0$-connections and $1$-connections also called  by {\it connective structure} and {\it curving}
respectively in \cite{Br1,Br2}. In \cite{Hit} the author does not separate
the notion of connection to 0 and 1 component, he calls by connection
the both structures together.

\subsection{Courant algebroids} \label{Cour-def}

Roughly speaking, Courant algebroid is a bundle
with the bracket on it, but the usual
conditions on the bracket like antisymmetricity, Leibniz rule,
Jacobi identity are relaxed in a certain,
very special way.

The term Courant algebroid originating from the
work \cite{Cour}
was proposed in \cite{LWX}
where  the main properties of Courant's bracket  (see formula
\ref{Cour-br1})
were axiomatized to the notion of Courant algebroid.
(In this paper it was shown that
the double of Lie bialgebroid is naturally a Courant algebroid.
One can see that naive wish to make a Lie algebroid
from the Lie bialgebroid fails and one
should use Courant algebroid instead.)

We refer to \cite{Roy} and references there in
 for comprehensive study
of Courant algebroid.

{\it Definition:} {\it Courant algebroid}
is a vector bundle $E$ equipped with a
nondegenerate symmetric bilinear form $\langle , \rangle$,
a bilinear bracket $[,]$ on $\Gamma (E)$, and
a bundle map $p : E\to TM$ satisfying the following properties:

\begin{enumerate}
\item   The left Jacobi identity
$[e_1 , [ e_2,  e_3]] =[ [e_1 e_2], e_3] + [e_2, [e_1, e_2]]$
\item Anchor is homomorphism $p [e_1, e_2] =[p (e_1), p(e_2)]$
\item Leibniz rule $[ e_1, fe_2]=f[ e_1, e_2] +\cL_{p (e_1)}(f) e_2 $
\item $[e, e] =\frac{1}{2}\cD \langle e,e\rangle$
\item Self-adjointness $p (e_1)\langle e_2,e_3\rangle =
\langle [e_1,  e_2] , e_3\rangle
+\langle e_2, [e_1,  e_3] \rangle $
\end{enumerate}
where $\cD$ is defined as
$p^* d : C^{\infty}(M)\stackrel{d}{\to} \Omega^1 (M)\stackrel{p^*}{\to}
 E^* \simeq E$.

One can also reformulate the definition of Courant
algebroid in terms of antisymmetric bracket:
$$\begin{array}{c}
[e_1,e_2]^{antisym}=\frac{1}{2}([e_1, e_2] - [e_2, e_1]) \\ { ~ }
[ e_1, e_1]=[e_1,e_2]^{antisym}+\frac{1}{2}\cD\langle e_1,e_2\rangle
\end{array}$$

Actually this was the point of view accepted in \cite{LWX},
but it is quite easy to reformulate one way to another.


\section{"Conducting bundle" construction of the Courant algebroid.} \label{Cond}
In this section we describe "conducting bundle" construction
following ideas of P. Severa. This construction
associates Courant algebroid to any extension
of bundles of the type $0 @>>> \cO @>>> \widetilde\cT @>{\sigma}>> \cT @>>> 0$
in canonical and functorial way. The main property of this
construction is
that, roughly speaking, the Lie-algebroid  structure on such extension
gives canonical trivialization of the obtained Courant algebroid.
More precisely we require not the arbitrary Lie algebroid structure
on this extension, but impose  some conditions
which are abstracted from the properties of Atiyah algebroids of the
line bundles. We call such extensions Picard-Lie-algebroids.
This construction can be used to attach Courant algebroid to the gerbe
(this will be described in the next section).
The intermediate step of the construction is to introduce
the "conducting bundle" $\cB_{\widetilde\cT}$,
let us remark that the term "bundle" is a bit misleading because
 $\cB_{\widetilde\cT}$
does not posses any structure of the $\cO$-module.

Let us briefly describe the constructions:
first step we introduce $\cB_{\widetilde\cT}$
as sheaf  endomorphisms of the extension  $0 @>>> \cO @>>>
\widetilde\cT @>{\sigma}>> \cT @>>> 0$ satisfying the
certain properties such as Leibniz rule, preservation the
extension structure and some other (see subsection \ref{def-B}),
the second step is to construct the Courant
algebroid as the factor of the fibered product
 $\cB_{\widetilde\cT}\times_\cT\widetilde\cT$
(see subsection \ref{con-Cour}).

The properties of the construction are
roughly speaking the following: $\cB_{\widetilde\cT}$
is an extension of the type:
$
0 @>>> \Omega^1 @>i>> \cB_{\widetilde\cT} @>{\pi}>> \cT @>>> 0
$ and  it has natural Lie algebra structure;
the constructed Courant algebroid is also
an extension of the type:
$
0 @>>> \Omega^1 @>>> \cQ_{\widetilde\cT} @>{\pi_\cQ}>> \cT @>>> 0
$, the bracket and scalar product  inherited
from the  $\cB_{\widetilde\cT}\times_\cT\widetilde\cT$
leads  precisely to the structure satisfying all axioms
for the Courant algebroid.

\subsection{Conducting Bundle}

\subsubsection{Definition of the conducting bundle.}
\label{def-B}\label{properties2}
For any extension of $\cO$-modules
\bea
0 @>>> \cO @>>> \widetilde\cT @>{\sigma}>> \cT @>>> 0
\eea
one has the sheaf
$\cB_{\widetilde\cT}$
(called {\it conducting bundle})
of $\bbC$-linear
endomorphisms of $\widetilde\cT$ which satisfy properties:
\begin{enumerate}
\item it preserves the inclusion $\cO @>>> \widetilde\cT$,
and acts on  $\cO$ as differentiation, hence
there is a map
\[
\pi : \cB_{\widetilde\cT} @>>> \cT
\]
which associates to a section of $\cB_{\widetilde\cT}$ it's action on $\cO$

\item the endomorphism, which  it induces on
 $\cT$ coincides with the action
of the vector field $\pi (t)$ by Lie derivative;
\item it satisfies the Leibnitz rule (i.e. $ b(fs)=\pi(b)(f)s+f b(s)$;
where $ b\in \cB;$ $ f\in \cO;$ $ s\in \widetilde\cT$,
in particular it is a differential operator of order one).
\end{enumerate}

{\it Remark:}
 $\cB_{\widetilde\cT}$
is not an $\cO$-module, i.e. the natural structure of
 $\cO$-module on $\bbC$-linear
endomorphisms of $\widetilde\cT$ does not
respect the subspace  $\cB_{\widetilde\cT}$.

\subsubsection{Motivation for the definition of the conducting bundle}
\label{Motiv}
Suppose that
\[
0 @>>> \cO @>>> \widetilde\cT @>{\sigma}>> \cT @>>> 0
\]
is Atiyah algebroid of some linear bundle. For each $t\in\widetilde\cT$ the endomorphism
\[
\ad(t) : s \mapsto [t,s]
\]
has the following properties:

\label{properties}
\begin{enumerate}
\item it preserves the inclusion $\cO @>>> \widetilde\cT$
\item the endomorphism it induces on $\cO$ and $\cT$ coincides with the action
of the vector field $\sigma(t)$ by Lie derivative;

\item it satisfies the Leibniz rule (i.e. $\ad(t)(fs)=\sigma(t)(f)s+f\ad(t)(s)$;
in particular it is a differential operator of order one).
\end{enumerate}

So one can see that the properties which define
$\cB_{\widetilde\cT}$ are precisely the same as the properties
of the operators of the adjoint representation of the  Atiyah algebroid.
So $\cB_{\widetilde\cT}$ is  the
unification of the all possible operators which can be represented
as adjoint representation for the Atiyah Lie-algebroids.

\subsubsection{Inclusion of 1-forms in  the conducting bundle.} \label{OmAct}

{~}

{\it Lemma:}
The natural action of $\Omega^1$ on $\widetilde\cT$ given by
$\alpha(t) = -\iota_{\sigma(t)}\alpha$ satisfies properties described
in the section \ref{properties2}. Hence, there
is a natural map
\[
\Omega^1  @>{i}>> \cB_{\widetilde\cT}\ .
\]

\subsubsection{The structure and the properties of the conducting bundle.}
\label{lem-B}

{~}

{\it Lemma:}
\begin{enumerate}
\item
The sequence
\[
0 @>>> \Omega^1 @>i>> \cB_{\widetilde\cT} @>{\pi}>> \cT @>>> 0
\]
is exact.
\item $\cB_{\widetilde\cT}$ is closed under the commutator bracket (of
endomorphisms of $\widetilde\cT$), hence a sheaf of Lie algebras.
\item The map $\pi$ is a morphism of sheaves of Lie algebras.
\item The inclusion of $\Omega^1$ is a morphism of sheaves of Lie
  algebras,
where we consider  $\Omega^1$ as an abelian Lie algebra.
\item
The bracket of the elements $b \in \cB_{\widetilde\cT}$ and
$ \alpha \in \Omega^1 \subset  \cB_{\widetilde\cT}$
takes the form:
\[
[b,\alpha]= + \cL_{\pi (b)} \alpha
\]

{\it Comment:} Sign in the formula above does not
depend on whether we define an action of $\Omega^1$
on  $\widetilde\cT$  by
$\alpha(t) = -\iota_{\sigma(t)}\alpha$
or by $\alpha(t) = + \iota_{\sigma(t)}\alpha$

\end{enumerate}

\subsubsection{Local description of conducting bundle.}

{\it Remark:} Locally, in terms of any isomorphism $\widetilde\cT = \cO\oplus\cT$,
the elements of $\cB_{\widetilde\cT}$
 are of the form:
\[
\left(
\begin{array}{cc} \xi & \alpha \\ 0 & \ad(\xi) \end{array}
\right)
\]
with $\xi\in\cT$ and $\alpha\in\Omega^1$.



\subsubsection{Automorphisms of extensions. }
The category of extensions of  $\cO$-modules of the form
\[
0 @>>> \cO @>>> \widetilde\cT @>{\sigma}>> \cT @>>> 0
\]
is a groupoid. The sheaf of local automorphism of an object is canonically isomorphic to
$\Omega^1$ with $\alpha\in\Omega^1$ acting by $t\mapsto t - \iota_{\sigma(t)}\alpha$.

\subsubsection{Automorphisms of conducting bundle.}
The category of extensions of the form
\[
0 @>>> \Omega^1 @>>> \cB @>{\pi_\cB}>> \cT @>>> 0
\]
is a groupoid. The sheaf of local automorphisms of an object is canonically isomorphic to
$\shHom(\cT,\Omega^1)$ with $\Phi\in\shHom(\cT,\Omega^1)$ acting by
$b\mapsto b + \Phi(\pi_\cB(b))$.

The local automorphisms which respect the Lie-algebra extension
 structure
 on
$\cB$ are canonically isomorphic
 morphisms
 $\shHom(\cT,\Omega^1)$ which are $d^{Lie}$-closed
i.e. are 1-cocycles of the Lie algebra of vector fields with
coefficients in  $\Omega^1$ ( this means that:
$
L_{\xi} \Phi (\eta )- L_{\eta}\Phi (\xi )=\Phi ([\xi ,\eta] )
$, note that this
does not hold for a $\cO-$linear $\Phi$).

\subsubsection{Functoriality of the conducting bundle.}
{~}

{\it Lemma:}
The association $\widetilde\cT \mapsto \cB_{\widetilde\cT}$ is functorial. The induced
map $Aut(\widetilde\cT) @>>> Aut(\cB_{\widetilde\cT})$ is given by
$\alpha\mapsto( b \mapsto b + [\alpha, b]= b- \cL_{\pi (b)}   \alpha)$.
Note that  $\cL_{\pi(b)} (\alpha)$ is just $(d^{Lie}\alpha) (\pi(b))$,
hence the corresponding  automorphism  respects    the Lie-algebra extension
 structure
 on
$\cB$.



\subsection{Construction of the Courant algebroid.}\label{con-Cour}
For any extension $\widetilde\cT$ there is a diagram
\[
\begin{CD}
0 @>>> \Omega^1 @>>> \cB_{\widetilde\cT} @>>> \cT @>>> 0 \\
& & @A{\pr_\Omega}AA  @A{\pr_\cB}AA  @AAA & \\
0 @>>> \Omega^1\oplus\cO @>>> \cB_{\widetilde\cT}\times_\cT\widetilde\cT @>>> \cT @>>> 0 \\
&  & @V{\pr_\cO}VV  @V{\pr_{\widetilde\cT}}VV  @VVV & \\
0 @>>> \cO @>>> \widetilde\cT @>>> \cT @>>> 0
\end{CD}
\]

Here we denote by $\cB_{\widetilde\cT}\times_\cT\widetilde\cT$
the fibered product of $\cB_{\widetilde\cT}$   and  $\widetilde\cT$
with respect to projections on $\cT$, i.e. subset
of  $\cB_{\widetilde\cT}\times \widetilde\cT$
such that $\pi (b)=\sigma( \bar t)$.

Let us construct {\it the Courant algebroid}
 $\cQ_{\widetilde\cT}$.  We define it
as the extension obtained from the middle row
by push-out via the map $\Omega^1\oplus\cO @>>> \Omega^1$ given by
$(\alpha,f) \mapsto \alpha - df$,
i.e. we factorize  $\Omega^1\oplus\cO$ and
$\cB_{\widetilde\cT}\times_\cT\widetilde\cT$  by the elements
of the form $ (df, f)$.
 Thus, $\cQ_{\widetilde\cT}$ fits into
\[
0 @>>> \Omega^1 @>>> \cQ_{\widetilde\cT} @>{\pi_\cQ}>> \cT @>>> 0
\]

We show below how to introduce the bracket and the scalar product
on $\cQ_{\widetilde\cT}$ such that all axioms of Courant algebroid
holds true.

{\it Comment:} We impose $(df,f)=0$ (or $(df,0)=(0,-f)$)
with this sign (but not  $(df,- f)=0$)
in order to have commutation relation
$ [\alpha, q]= (-\cL_{\pi(q)}\alpha+\iota_{\pi(q)}(\alpha))
$.

The signs should be agreed in the following two
places:

1) The natural action of $\Omega^1$ on $\widetilde\cT$ given by
$\alpha(t) = -\iota_{\sigma(t)}\alpha$ (see section \ref{OmAct})

2) $(df,f)=0$

The result of the agreement is that we have right
commutation relation:

$ [\alpha, q]= (-\cL_{\pi(q)}\alpha+\iota_{\pi(q)}d(\alpha))
=-d \iota_{\pi(q)} \alpha $

(see formula \ref{F})

\subsubsection{The bracket on $\cB_{\widetilde\cT}\times_\cT\widetilde\cT$.}
The sheaf $\cB_{\widetilde\cT}\times_\cT\widetilde\cT$ is naturally endowed with the
bi-linear operation (refered to as ``bracket'') defined by the formula
\[
[(b_1,t_1),(b_2,t_2)] = ([b_1,b_2], b_1(t_2))\ .
\]

{\it Remark:} note that the usual semidirect product Lie bracket
$[(b_1,t_1),(b_2,t_2)] = ([b_1,b_2], b_1(t_2)-b_2(t_1) )$ obviously
does not preserve the subspace
$\cB_{\widetilde\cT}\times_\cT\widetilde\cT$.
But described above not antisymmetric bracket is well-defined on this
subspace.

{\it Lemma:}
This operation satisfies the left Jacobi identity (i.e. the
left  adjoint action is by derivations).

This is obvious because $\cB_{\widetilde\cT}$ is a Lie algebra and $\widetilde\cT$ is a
$\cB_{\widetilde\cT}$-module. The bracket is  not skew symmetric, and, thus,
$\cB_{\widetilde\cT}\times_\cT\widetilde\cT$ is only a Leibniz algebra
\cite{Lod}.

\subsubsection{The bracket on $\cQ_{\widetilde\cT}$.}\label{Cour-brac}
{~}

{\it Lemma:}
The bracket on $\cB_{\widetilde\cT}\times_\cT\widetilde\cT$ descends
 to $\cQ_{\widetilde\cT}$, i.e. the subspace $(df,f)$
is the left and right ideal for the bracket on
$\cB_{\widetilde\cT}\times_\cT\widetilde\cT$,
(it is obviously not true for the hole space
$\cB_{\widetilde\cT}\times \widetilde\cT$).
For $q_i \in  \cQ_{\widetilde\cT}$ and
 $\alpha_i \in \Omega^1 \subset \cQ_{\widetilde\cT}$
the bracket takes the form:

\[
[q_1+\alpha_1,q_2+\alpha_2]=
 [q_1,q_2] + \cL_{\pi_{\cQ}(q_1)}\alpha_2-
\cL_{\pi_{\cQ}(q_2)}\alpha_1
+ d ( \iota_{\pi_{\cQ}(q_2)}\alpha_1)
\]
or it can be rewritten in the form
\[
[q_1+\alpha_1,q_2+\alpha_2]=
 [q_1,q_2] + \cL_{\pi_{\cQ}(q_1)}\alpha_2
-  \iota_{\pi_{\cQ}(q_2)} d (\alpha_1)
\]

So, after the  choice of splitting and antisymmetrization
the bracket  obviously coincides with the
original Courant's bracket \ref{Cour-br1}.

The proofs are straightforward so let us only comment on
 the appearance of the additional
term
 $ d ( \iota_{\pi_{\cQ}(q_1)}\alpha_2)$
in the above formula.
It easy to see from the definitions
and item 5 of lemma \ref{lem-B}
that for the $ (\alpha,0) \in
\cB_{\widetilde\cT}\times_\cT\widetilde\cT$,
where  $\alpha \in \Omega^1 \subset\cB_{\widetilde\cT}$
and $(b,\xi) \in\cB_{\widetilde\cT}\times_\cT\widetilde\cT$
the bracket takes the form

\[
[ (\alpha,0),(b,\xi)  ]=
(-\cL_{\pi(b)}\alpha, - \iota_{\sigma(\xi)}(\alpha))
\]

In  $\cQ_{\widetilde\cT}$ the elements $ (df,0)$ and $(0,- f)$
are identified, hence the formula above  takes the form:

\begin{eqnarray}
[ \alpha,b  ]=
-\cL_{\pi(b)}\alpha+ d \iota_{\pi(b)}(\alpha))
= - \iota_{\pi(b)}( d\alpha) \label{F}
\end{eqnarray}

where we have also used that ${\pi(b)}={\sigma(\xi)}$.

{\it Remark:}
the second way to write the formula above
emphasize the $\cO$-linearity in second argument.

{\it Remark:}
Note that

\[
[ (b,\xi),(\alpha,0)  ]=
(\cL_{\pi(b)}\alpha, 0)
\]

without any additional terms.

\subsubsection{The pairing on
$\cB_{\widetilde\cT}\times_\cT\widetilde\cT$.} \label{Cour-pair}
The symmetrized bracket
\[
[(b_1,t_1),(b_2,t_2)] + [(b_2,t_2),(b_1,t_1)] = (0, b_1(t_2) + b_2(t_1))
\]
leads to the symmetric operation
\begin{eqnarray*}
<\ ,\ > : \cB_{\widetilde\cT}\times_\cT\widetilde\cT\times
\cB_{\widetilde\cT}\times_\cT\widetilde\cT  & @>>> & \cO \\
((b_1,t_1),(b_2,t_2)) & \mapsto & (b_1(t_2) + b_2(t_1))
\end{eqnarray*}


\subsubsection{The pairing on  $\cQ_{\widetilde\cT}$ and its properties}
\label{Cour-pair-prop}
{~}

{\it Lemma:}
\begin{enumerate}
\item $\left(\cB_{\widetilde\cT}\times_\cT\widetilde\cT\right)^\perp = \{(df,f)\vert f\in\cO\}$
\item $\Omega^1\perp\Omega^1$
\item The pairing descends to a non-degenerate pairing on $\cQ_{\widetilde\cT}$, i.e.
the map
\begin{eqnarray*}
\cQ_{\widetilde\cT} & @>>> & \shHom(\cQ_{\widetilde\cT}, \cO) \\
q & \mapsto & <q,\bullet>
\end{eqnarray*}
is a monomorphism.
\item The diagram
\[
\begin{CD}
\cQ_{\widetilde\cT} @>>> \shHom(\cQ_{\widetilde\cT}, \cO) \\
@AAA @AA{\pi_\cQ}A \\
\Omega^1 @>>> \shHom_\cO(\cT,\cO)
\end{CD}
\]
is commutative, i.e.
$<\alpha , q > = \iota_{\pi(q)} \alpha$.

\end{enumerate}

\subsubsection{Corollary}\label{Cour-pair-prop2}
\begin{enumerate}
\item The induced bracket on $\cQ_{\widetilde\cT}$ satisfies
\[
[q_1,q_2] + [q_2,q_1] = d<q_1,q_2> \ .
\]
\item The inclusion $\Omega^1 @>>> \cQ_{\widetilde\cT}$ is the adjoint of the projection
$\cQ_{\widetilde\cT} @>>> \cT$.
\end{enumerate}

\subsubsection{Functoriality.}

{\it Lemma:} one can easily see that
the association $\widetilde\cT \mapsto \cQ_{\widetilde\cT}$ is functorial. The induced
map $Aut(\widetilde\cT) @>>> Aut(\cQ_{\widetilde\cT})$ is given by
$\alpha\mapsto( q \mapsto q - \iota_{\pi_{\cQ}(q)} d\alpha)$.

\subsubsection{$\cQ_{\widetilde\cT}$ is Courant algebroid.}
\label{Theor1}
{~}

{\it Theorem:} $\cQ_{\widetilde\cT}$ with bracket
and the pairing  defined above, and the $\cO$-module structure
which will be defined latter is Courant algebroid.

In previous sections we have already established
that the bracket and the pairing satisfy all the necessary axioms
of Courant algebroids. The only thing to establish is the
$\cal O$-module structure and prove Leibniz rule for it.
This will be done below (see section \ref{ExCour}).
At the moment we do not know natural construction
for the $\cal O$-module structure, the reason is that
conducting bundle $\cB_{\widetilde\cT}$
does not have  an $\cal O$-module structure.
So in the next section we will  introduce the
$\cal O$-module structure in terms of some splitting (called connections
there)
and we will check that it is independent of the choice of splitting.

\subsubsection{Remark:}
in the
smooth case any Courant algebroid obtained by such construction
is isomorphic to the standard one, non trivial examples can be obtained from gerbes
(see section \ref{gerb}).


\subsection{Picard-Lie algebroids and trivialization of
Courant algebroids constructed from the Picard-Lie algebroids.}

The idea of this subsection is, roughly speaking, that
Courant algebroids constructed from the extension
with brackets are canonically trivialized.
More precisely one needs not the arbitrary bracket,
but the one with the properties similar to the
properties of the bracket on Atiyah algebroids. Such extensions we will call
Picard-Lie algebroids.

\subsubsection{Picard-Lie algebroids.}\label{Pic-Lie}
Let us introduce the  notion which we call Picard-Lie algebroid,
roughly speaking it is Lie algebroid
of the type:
$0 @>>> \cO @>>> \widetilde\cT @>{\sigma}>> \cT @>>> 0$,
with some additional requirement. This definition is just
abstractization of the properties of the Atiyah algebroid
of some line bundle, relaxing the only requirement
of integrality of periods for the Chern class of linear bundle.

{\it Definition:} The extension
$0 @>>> \cO @>>> \widetilde\cT @>{\sigma}>> \cT @>>> 0$
is called {\it Picard-Lie algebroid} (PLA) if:
\begin{enumerate}
\item $ \widetilde\cT$ is Lie algebroid with anchor $\sigma$;
\item $\cO$ is abelian ideal in $\widetilde\cT$;
\item $[\bar t, f]=\cL_{\sigma(\bar t)} f$
for $f \in \cO \subset\widetilde\cT$.
\end{enumerate}

{\it Example:} Atiyah algebroid of any linear bundle
is Picard-Lie algebroid.

{\it Remark:}
obviously all the properties from the  section \ref{Motiv}
holds true for the endomorphisms $ad_{\bar t}$ for any
Picard-Lie algebroid.


\subsubsection{The construction of the splitting of Courant algebroid
coming from PLA} \label{spl}
Suppose that $\widetilde\cT$ is, in fact, a PLA. The adjoint representation of $\widetilde\cT$
gives rise to the splitting of $\pr_{\widetilde\cT}$, namely
$\bar t\mapsto(\ad(\bar t),\bar t)$.

{\it Lemma:} on the subspace $\cO \subset \bar \cT$
this splitting restricts
to the splitting $f\mapsto(df,f)$ of $\pr_\cO$.

(Here we essentially used that $\cT$ is indeed PLA and not arbitrary
Lie algebroid).

{\it Construction of splitting:} consider $t\in \cT$
and take $\bar t \in \bar \cT$
such that $\sigma (\bar t)=t$,
define the splitting $A: \cT \to \cQ_{\widetilde\cT}$ by the
formula $ t \mapsto (\ad(\bar t),\bar t) \in \cQ_{\widetilde\cT}$.

{\it Lemma:} the splitting above does not depend on the choice of
$\bar t$.

{\it Proof:} the different choices of
$\bar t$ differs by some $f\in \cO$. Due to the previous
lemma $f\mapsto(df,f)$, by the construction  the Courant algebroid
is factorization of $ \cB_{\widetilde \cT}\times_{\cT} \widetilde \cT$
by the elements of the form $(df,f)$, hence the splitting is well-defined.

\subsubsection{Properties of the splitting}

{\it Lemma:} the following holds true:
\begin{enumerate}
\item the image of the splitting $\cT \to \cQ_{\widetilde\cT}$
is isotropic with respect to the pairing on  $\cQ_{\widetilde\cT}$;
\item the image of the splitting $\cT \to \cQ_{\widetilde\cT}$
is closed with respect to the bracket on $\cQ_{\widetilde\cT}$;
\item the splitting  $A: \cT \to \cQ_{\widetilde\cT}$
is a homomorphism of Lie algebra to the Leibniz algebra.
\end{enumerate}

As a corollary of the all above we obtain.

{\it Theorem:}
Suppose that $\widetilde\cT$ is a PLA then,
the map $ \Omega^1 \oplus \cT \to \cQ_{\widetilde\cT}$
given by $(\alpha, t) \mapsto (\alpha, A(t))$ is an isomorphism
of the standard  Courant algebroid  $ \Omega^1 \oplus \cT$
and the Courant algebroid $\cQ_{\widetilde\cT}$.


\maketitle
\section{Exact Courant} \label{exCour}

In this section following P. Severa (see also
\cite{Roy} pages 48-50) we show that
one can build a kind of differential geometry for
Courant algebroids, one can introduce the notion
of connection and  its curvature.
More precisely this can be done for the Courant
algebroids of the  type:
$ 0 @>>> \Omega^1 @>>> \cQ @>{\pi_\cQ}>> \cT @>>> 0$
which are called {\it exact} Courant algebroids.
But in contrast to the usual situation
the curvature form is closed 3-form (not the 2-form as usually)
the difference of two connections is 2-form (not the 1-form as usually).
Further Severa remarked that, in analogous to the
usual situation, adding some 2-form $\omega$ to some
connection one sees that the curvature 3-form changes to the $d \omega$,
hence we have the well-defined characteristic class
of the Courant algebroid (Severa's class).
On the other hand starting from the standard Courant
algebroid and some closed 3-form Severa
proposed to twist the Courant algebroid
and to obtain new Courant algebroid.

So one can see that the theory of Courant algebroids
is in a sense analogous to the theory of gerbes.
This is not accidently and following ideas of P.Severa
we show how to attach the Courant algebroid to the gerbe.

In the last subsection we remark that WZNW-Poisson condition
introduced recently by Klimcik-Strobl is
the same as the condition for graph of the bivector $\pi$
to be   closed under the Courant's bracket in the
twisted Courant algebroid constructed by the closed
3-form $H$.

At the beginning of the section we pay the debt
from the previous section
and finish the prove of theorem from the section
\ref{Theor1} that the conducting bundle really gives
the Courant algebroid, the only thing to do is to introduce
the $\cO$-module structure on it and to prove the Leibniz rule
for it. We do this in this section, because to prove
it is more convenient to introduce the notion of connection
and to develop some of its properties,
after  doing this the proof follows easily.

\subsection{Exact Courant algebroids.}
\subsubsection{\it Definition.}
Let us call Courant algebroid  $\cQ$  {\it exact} Courant algebroid
if the following holds true:
\begin{enumerate}
\item
 $0 @>>> \Omega^1 @>>> \cQ @>{\pi_\cQ}>> \cT @>>> 0$,
where ${\pi_\cQ}$ is anchor map;
\item $ \Omega^1$ is an abelian ideal in $\cQ$;
\item the pairing satisfy: $<\alpha, q > = \iota_{\pi(q)} \alpha $.
\end{enumerate}

{\it Remark:} the Courant algebroid obtained by the
conducting bundle construction is obviously exact.
Below we will see that up to isomorphism the exact Courant
algebroids can be classified by the $H^3(M,\bbR)$.

\subsubsection{Alternative description of exact Courant algebroid.}
 \label{ExCour}
In this section we will give slightly different
description of exact Courant algebroid, actually
the main thing to prove is that $\cO$-module structure follows from  the
other properties. The reason to do this is to pay the debt from the
previous section - see corollary below.

{\it Proposition:}
The sheaf of vector spaces $\cQ$ such that the
properties below holds true
is  exact Courant algebroid:
\begin{enumerate}
\item there is an exact sequence:
\[
0 @>>> \Omega^1 @>>> \cQ @>{\pi_\cQ}>> \cT @>>> 0
\]

\item there is an
bi-linear operation $[,]$ (refered to as ``bracket'')
which satisfies left Jacobi identity (i.e. left adjoint
action is by differentiation)
\item there is an symmetric bi-linear nondegenerate  pairing $<,>:
\cQ  \times  \cQ\,
@>>>
\, \cO $
\item the maps in the sequence
$
0 @>>> \Omega^1 @>>> \cQ @>{\pi_\cQ}>> \cT @>>> 0
$ respect the corresponding brackets (we consider $\Omega ^1$
as the abelian Lie algebra)

\item
the operations satisfy the following compatibility
condition:
\begin{eqnarray}
[q_1,q_2] + [q_2,q_1] = d<q_1,q_2> \label{dP} .
\end{eqnarray}

\item The pairing is invariant i.e.

\begin{eqnarray}
<[q_1,q_2],q_3>+   <q_2,[q_1,q_3]>=\cL_{\pi_{\cQ}(q_1)}<q_2,q_3>
\label{pair-invar}
 \end{eqnarray}



\item
the pairing between the
$ \alpha \in \Omega^1 \subset  \cQ$ and $q \in \cQ$
is given by the formula
\begin{eqnarray}
<\alpha,q>= \iota_{{\pi_\cQ}(q)} \alpha \label{def-pair}
 \end{eqnarray}

\item for $q \in \cQ$ and $\alpha \in \Omega ^1 \subset \cQ$
the bracket is given by the formula:

\begin{eqnarray}
[q,\alpha]=\cL_{\pi_{\cQ}(q)} \alpha  .
\end{eqnarray}

\end{enumerate}

{\it Corollary:} this proposition finishes the proof
of the main theorem from the previous section (see section
\ref{Theor1}, that the conducting bundle construction really gives
the Courant algebroid).

The corollary is obvious because we have already established
that all the
requested in the proposition above properties
   holds true
for the
result of the conducting bundle construction,
(see lemmas in the sections \ref{Cour-brac},\ref{Cour-pair},
\ref{Cour-pair-prop2})
 hence there exist
$\cO$-module structure with the required properties, hence
it is Courant algebroid.

%
\subsubsection{}
The proposition above is actually equivalent to the following lemma.

{\it Lemma:}
Let $\cQ$ satisfy the properties of the proposition above
then:

\begin{enumerate}

\item
There is natural
 $\cO$-module structure on
 $\cQ$
and
the following holds:

\begin{eqnarray}
[ q_1,f q_2]=f[ q_1, q_2]+ ( \cL_{\pi_{\cQ}(q_1)} f  )  q_2
\label{Fun1} \\ {  ~ }
 [ f q_1, q_2]= f[ q_1, q_2]-( \cL_{\pi_{\cQ}(q_2)} f  )  q_1
+< q_1, q_2> df\label{Fun2} \\ {  ~ }
 < f    q_1, q_2>=  f < q_1, q_2> \label{Fun3}
 \end{eqnarray}

\item  $\Omega^1\perp\Omega^1$

\item

\[
[q_1+\alpha_1,q_2+\alpha_2]=
 [q_1,q_2] + \cL_{\pi_{\cQ}(q_1)}\alpha_2-
\cL_{\pi_{\cQ}(q_2)}\alpha_1
+ d ( \iota_{\pi_{\cQ}(q_2)}\alpha_1)
\]

\item
in the splitted situation $\cQ=\cT \oplus \Omega^1$
the pairing is given by the formula
\[
<(\xi_1, \alpha_1),(\xi_2, \alpha_2)>=
\iota_{\xi_1}( \alpha_2 )+\iota_{\xi_2}( \alpha_1 )
\]

\end{enumerate}

In order to see that the properties above  holds true
we will do, roughly speaking, the following:
we introduce the isotropic splitting of the ${\pi_{\cQ}(q)} $
( which will be  called {\it  connection})
and prove everything with it help,
after it can be seen
that all the properties does not depend on the choice of connection.
To do this some properties of connections should be mentioned.
The $\cO$-module structure will be introduced in the
section \ref{O-mod} and  the proof of the properties
will be finished in section \ref{Proof-O-mod}.


\subsection{Connection and its  curvature for the Courant algebroid.}
\subsubsection{Definition.}

Let us  call the {\it connection} on  ${\cQ}$
the  map  $A: \cT @>>> {\cQ}$, such that
it splits projection $\pi_{\cQ}$ i.e.  $\pi_{\cQ} A= id$,
(hence the image of $A$
is
 transversal to the
$\Omega^1 \subset {\cQ}$)  and    isotropic
(i.e. $< A(\xi_1),A(\xi_2) > =0$).

\subsubsection{Remark}
The space of connections is affine space
under the vector space of $\Omega^2$.
The connection $ A+\omega$
is given by the map $A+\omega:
\xi \mapsto A(\xi)+\iota_\xi \omega$.
One can easily see from the formula \ref{def-pair}
that the new map will also be isotropic
 and it's obviously splits the map $\pi_{\cQ}$ i.e.  $\pi_{\cQ} (A+\omega)= id$.

\subsubsection{Lemma:} \label{ConLem}
All connections can be obtained from the given
one adding to it  some 2-form $\omega$.

Consider two connections $A_1$ and $A_2$.
Let $\alpha_\xi= A_1(\xi)-A_2(\xi)$.
From transversality follows that
  $\alpha_\xi$ is 1-form
for any  given $\xi$.
Using isotropicity one obtains:
$
0=<A_1(\xi_1)+ \alpha_{\xi_1},
A_1(\xi_2)+ \alpha_{\xi_2}>$
hence
$\iota_{\xi_2} \alpha_{\xi_1}
= \iota_{\xi_1} \alpha_{\xi_2}$
hence
 $\alpha_\xi=\iota_\xi \omega$
for some 2-form $\omega$.

\subsubsection{Construction of the $\cO$-module structure on ${\cQ}$.}
\label{O-mod}
Let us choose any connection $A$.
Any element $q \in {\cQ}$ can
be represented uniquely as $q= \alpha + A(\xi)$, for
some $\alpha$ and $\xi$.
Let us define
\[
f q \overset{def}{=}A(f\xi)+f \alpha
\]

Obviously such multiplication by functions
commutes with the change of connection given
by $ A \to A+ \omega$ i.e.
$f(( A+\omega)(\xi))=(A+\omega)(f\xi)$
hence
in  view of lemma \ref{ConLem}
the multiplication by functions does not
depend on the choice of connection.


\subsubsection{Definition:}
let us call the  {\it curvature}
of the connection $A$
the map $F: \cT\times\cT @>>> \Omega^1$
given by $F(x,y)=[A(x),A(y)]-A([x,y])$.
Due to the isotropicity of connection
and the formula \ref{dP} this map is antisymmetric
in $x$ and $y$.

Let us call the
 3-form $H(\xi_1,\xi_2,\xi_3)$
defined by the formula
$H(\xi_1,\xi_2,\xi_3)= \iota_{\xi_3} F(\xi_1,\xi_2)$
 the {\it  curvature 3-form} for
the connection $A$.

\subsubsection{Lemma}
One can obviously see from the isotropicity
of connection and formula \ref{def-pair}
that
\begin{eqnarray}
H(\xi_1,\xi_2,\xi_3)=< [A(\xi_1),A(\xi_2)],A(\xi_3)>
\label{Curv}
\end{eqnarray}

\subsubsection{Lemma}\label{Curv-prop}
Curvature 3-form satisfies the following properties:
it is antisymmetric in all three arguments,
it is $\cO$-linear and it is closed.

To argue  these properties
one goes as follows.
By the invariance of the pairing
( formula \ref{pair-invar})
and isotropicity of connection
one has:
$< [A(\xi_1),A(\xi_2)],A(\xi_3)>= -
 < [A(\xi_1),A(\xi_3)],A(\xi_2)>$.
Hence curvature 3-form is antisymmetric
in all three arguments.

Is is obviously $\cO$-linear in
the third argument,
hence by antisymmetricity it
is  $\cO$-linear in all three
arguments.

Closedness (Bianchi identity) follows from the Jacobi identity
as usually :

\[
Jac ( A(\xi_1),A(\xi_2),A(\xi_3))
=  - \iota_{\xi_1 }\iota_{\xi_2 }\iota_{\xi_3 }
d H
\]

where $Jac(q_1,q_2,q_3)$ is "Jacobiator"
given by the formula $ [q_1,[q_2,q_3] ]
-[[q_1,q_2]   ,q_3]-
[q_2,[q_1,q_3] ]
$

To check this formula it is convenient to use
the formula \ref{K2}, it is obvious that the above
reasoning is analogous to the one in section \ref{Deform}.

\subsubsection{Proof of the formulas
\ref{Fun1},\ref{Fun2},\ref{Fun3}.} \label{Proof-O-mod}
The formula \ref{Fun3} obviously follows from
the formula \ref{def-pair} and isotropicity of connection.
To prove \ref{Fun1} one should note that
it's  holds for the case $[q, f \alpha]$
(in view of the  formula \ref{def-pair}).
To check it for $[ A(\xi_1),f A(\xi_2)]$
one proceeds as follows:
 $[ A(\xi_1),f A(\xi_2)]=
A[\xi_1,f \xi_2 ] + F(\xi_1,f \xi_2)
= f A[\xi_1, \xi_2 ]+ (\cL_{\xi_1}f)A(\xi_2)+
f F(\xi_1,\xi_2)
= f[ A(\xi_1), A(\xi_2)] +(\cL_{\xi_1}f)A(\xi_2)
$.
We have used the
 $\cO$-linearity of the curvature $F$.

So the formula
\ref{Fun1} is proved.
The formula \ref{Fun2}
follows from it and the formula \ref{dP}.

%

\subsection{Further properties of exact Courant algebroids.}

We will study isomorphisms, deformations and classification of
 exact Courant algebroids.

\subsubsection{Isomorphisms}

{\it Lemma:}
The sheaf of local isomorphisms
of ${\cQ}$ is canonically isomorphic to the sheaf
of closed 2-forms $ \Omega^{2,cl}$
the action is given by the formula
$\Phi_\omega: ~ q \mapsto q+ \iota_{\pi_{\cQ}(q)}\omega$.

This can be seen from the formula:
\begin{eqnarray}\label{Iso1}
[\Phi_\omega (q_1),\Phi_\omega (q_2) ]=
\Phi_\omega  [q_1, q_2] -
\iota_{\pi_{\cQ}(q_1)}\iota_{\pi_{\cQ}(q_2)}
d \omega
\end{eqnarray}

This formula follows from \ref{K1}.

\subsubsection{Twisting the Courant algebroid by the closed 3-form.}
\label{Deform}

{~}

{\it Lemma:}
The sheaf of local deformations
of the bracket on ${\cQ}$ is canonically isomorphic to the sheaf
of closed 3-forms $ \Omega^{3,cl}$
the corresponding deformation  is given by the formula:
$ [q_1,q_2]_{new}= [q_1,q_2]+ \iota_{ \pi_{\cQ}(q_1)  }
 \iota_{ \pi_{\cQ}(q_2)  } H$.

This can be seen from the formula:

\[
Jac ( q_1, q_2 , q_3)_{new}=
 \iota_{ \pi_{\cQ}(q_1)  }
 \iota_{ \pi_{\cQ}(q_2)  } \iota_{ \pi_{\cQ}(q_3)  }
d  H
\]

Where $Jac ( q_1, q_2 , q_3)_{new}$
is "Jacobiator" with respect to the new bracket,
i.e.
 $Jac ( q_1, q_2 , q_3)_{new}=
[q_1,[q_2,q_3]_{new} ]_{new}
-[[q_1,q_2]_{new}   ,q_3]_{new}-
[q_2,[q_1,q_3]_{new} ]_{new}
$

Note that such deformations preserve the
pairing.

To check this formula it is convenient to use
the formula \ref{K2}, it is obvious that the above
reasoning is analogous to the one in section \ref{Curv-prop}.

\subsubsection{Lemma}
The deformations which correspond to exact 3-forms
$H= d \omega$
are trivial, the isomorphism
is given by the expected rule:
$q \mapsto q+ \iota_{\pi_{\cQ}(q)}\omega$.
This can be seen from the formula \ref{Iso1}

\subsubsection{Lemma}
The curvature of the connection
$ A+\omega$ equals to $H(A)+ d\omega $.

This statement  is essentially equivalent
to the formula \ref{Iso1}.


\subsubsection{Characteristic class of exact Courant
  algebroid. (Severa's class).}
So we have seen that  taking
the curvature 3-form of  any connection
one obtains closed 3-form,
we have also seen that taking another connection
curvature changes up to exact 3-form.

So for the given exact Courant algebroid
there is well-defined cohomology class $ [H] \in H^3(M)$.

{\it Example:}
if we take standard
Courant algebroid and the
connection $\xi \mapsto (\xi,0) \in \cQ$,
then  the curvature of such connection is
zero.
If we take the Courant algebroid
obtained by deformation of standard one by
3-form $\Omega$ (see section \ref{Deform})
and the same connection
then it's curvature 3-form equals to $-\Omega$.

\subsubsection{Addition.}
For the extensions $\cQ_1$ and $\cQ_2$:
$
0 @>>> \Omega^1 @>>> \cQ_i @>{\pi}>> \cT @>>> 0
$
one can define their Baer sum. Let us remark that
if $\cQ_i$ are Courant algebroids then there is natural
Courant algebroid structure on the Baer sum of  $\cQ_i$.
The characteristic class of it equals to the sum of characteristic
classes.

To check this one proceeds as follows: consider
the fibered product of
$\bar \cQ  \overset{def}{=}  \cQ_1 \times_{\cT} \cQ_2  $
by definition the Baer sum $\cQ_{1+2}$ is factor of $\bar \cQ $
by the skewdiagonal $ ( \alpha, - \alpha )$ where
$\alpha \in \Omega ^1 \subset \cQ_i$,
with the embedding of $\Omega ^1 @>>> \cQ_{1+2}$
given by $\alpha \to (\alpha , 0) $.
So one has
$
0 @>>> \Omega^1 @>>> \cQ_{1+2} @>{\pi}>> \cT @>>> 0
$
.
 The bracket and the  pairing  on $\bar \cQ $
are induced from the imbedding  $\bar \cQ  \subset \cQ_1 \times \cQ_2
$,
one can easily see that $\bar \cQ$ is subalgebra in $\cQ_1 \times
\cQ_2$.
To endow with the bracket and pairing the $\cQ_{1+2}$
one easily checks that skewdiagonal $ ( \alpha, - \alpha )$
is both-sided ideal for the bracket and
orthogonal to
$\bar \cQ $.
So both the pairing and the bracket can be
defined on  $\cQ_{1+2}$.

\subsubsection{Generalization. (We are grateful for A. Kotov
for suggestive discussion on this point) }

All the statements of this section are straightforwardly
true for any Courant algebroid
of the form:

\[
0 @>>> \cA^*  @>>> \cQ @>{\pi_\cQ}>> \cA @>>> 0
\]

where $\cA$ is any algebroid
and $\cQ$ is defined by the axioms \ref{ExCour}.
We note that in this situation
there is natural map $ \Omega^1 @>>>
\cA^*$ which is dual to the anchor $\cA @>>> \cT$,
hence formulas like \ref{dP} make sense.
All propositions holds true in this situation
because all the formulas
of differential calculus:
Lie derivatives, deRham complex,
$\iota_{a}$ etc. holds identically  true for any
Lie algebroid $\cA$ not only
for $\cT$ (see \cite{Mac} or for example
section 2 of \cite{ELW}).
For example such Courant algebroids are classified
by the $H^3(M,\cA)$ - Lie algebroid cohomology
of $\cA$.


\subsection{Courant algebroid of gerbe.}\label{gerb}

Both DD-gerbes and exact Courant algebroids are classified by the third cohomology
group, so one can just take the class of the gerbe and construct the exact
Courant algebroid with the same class, but one can hope there is more direct
relation.
As we already mentioned
P. Severa proposed that the role of the Courant algebroid  is an Atiyah algebroid of a gerbe, but
at the moment not everything is clear with this analogy.
Let us sketch the construction (idea of which is due to Severa's)
of Courant algebroid from  gerbes.
Construction consists of three step.

Consider  gerbe $F$  on $M$.

{\bf Step 1.}
Gerbes are classified by $H^{3}(X, \mathbb{Z})$=$H^{2}(X,
\underline{\mathbb{C}^*})$, where following Brylinski we denoted by
$\underline{\mathbb{C}^*}$ the sheaf of invertible functions.
Let us choose the Cech representative of the
characteristic class of gerbe, it can be  given by some invertible function
$g_{\alpha \beta \gamma}$ on the intersection of charts $U_{\alpha \beta
\gamma}$.
Let us choose the 1-forms $A_{\alpha \beta}$ on each intersection of charts
$U_{\alpha \beta}$, such that they satisfy the condition:

\bea
A_{\alpha \beta}+A_{\beta \gamma} + A_{\gamma \alpha} = d log (g_{\alpha \beta
\gamma})
\eea

This can be always done for the small enough covering $U_{\alpha}$, because
$g_{\alpha \beta
\gamma}$ is Cech cocycle. As it is explained in \cite{Br2} such choice is
essentially  the choice of what is called "0-connection on gerbe" or "connective
structure".

{\bf Step 2.}
Let us consider on each chart $U_{\alpha}$ bundles $B_{\alpha}= {\mathbb{C}} U_{\alpha}
\oplus \cT U_{\alpha}$.
On the intersection of charts $U_{\alpha \beta}$ consider the maps
$ \Phi_{\alpha \beta}: B_{\alpha} \to B_{\beta}$ given by the 1-forms $A_{\alpha
\beta}$ by the rule: $ \Phi_{\alpha \beta}: (f, \xi) \mapsto (f+ \iota_{\xi}
A_{\alpha \beta}, \xi)$.
Obviously one cannot glue together $B_{\alpha}$ to the global object
because the triple product $\Phi_{\alpha \beta}\Phi_{\beta \gamma} \Phi_{\gamma
\alpha}$ is not identity, but given by the closed 1-form
$  d log (g_{\alpha \beta
\gamma})$.

{\bf Step 3.}
Let us apply the conducting bundle construction to each $B_{\alpha}$.
We obtain Courant algebroid $C(B_{\alpha})$ on each chart $U_\alpha$,
by the functoriality of the construction we have the maps  $ C(\Phi_{\alpha \beta}): C(B_{\alpha}) \to C(B_{\beta})$
on the intersection of charts $U_{\alpha \beta}$.

The main point is that the triple product
$C(\Phi_{\alpha \beta}) C(\Phi_{\beta \gamma}) C(\Phi_{\gamma
\alpha})$ does equals the identity, because the automorphism of extension $B_{\alpha}$ given
by the 1-form $\omega$ induces automorphism of the   Courant
algebroid given by 2-form $d \omega$, hence
$C(\Phi_{\alpha \beta}) C(\Phi_{\beta \gamma}) C(\Phi_{\gamma
\alpha})= Id + d (d log (g_{\alpha \beta
\gamma}))=Id $. Hence we have the globally defined exact Courant algebroid.
One can obviously see that its characteristic class coincides with
the characteristic class of the gerbe, more precisely with its image in
$H^3 ( \mathbb{Z}) \otimes  \mathbb{R}$.

The construction is finished.

Let us make few comments on the construction. First one can see that the
construction is not direct: we work with the characteristic class of gerbe,
but not with the sheaf of groupoids itself, it would be very interesting
if on can propose the construction analogous to Atiyah algebroid construction
which works with the sheaf of groupoid itself and makes precise sense
of the "groupoid of its symmetries" and "Courant algebroid of its infinitesimal
symmetries". Second P. Severa remarked that the choice of 1-connection on
gerbe leads to connection on Courant algebroid (in a sense explained in
section 4) this is in complete analogy with line bundle - Atiyah algebroid
situation, where the  connection is the same as the splitting of Atiyah
algebroid. Third  P. Severa remarked that bundles $B_{\alpha}$ considered
in step 3 of the construction should be understood as connective structure
functor (see \cite{Br1,Br2}) applied to the object of groupoid on $U_{\alpha}$
(there is only one object up to isomorphism for the small enough
$U_{\alpha}$).
The map $B_{\alpha}\to B_{\beta}$ should arise (after the choice of trivialization of
$A(L_{\alpha\beta})$)  from the map
$B_{\alpha}\to B_{\beta} \underset{Baer}{\oplus} A(L_{\alpha\beta})$
which should arise from the some functoriality
properties, where $L_{\alpha\beta}$ is the line bundle on the intersection of
charts which is part "gerbe data" description of gerbes.
This picture is very interesting because it works directly with the sheaf
of groupoids, not only with the characteristic classes of gerbes,
but at the moment not all the details clear for us how does the map
$B_{\alpha}\to B_{\beta}\underset{Baer}{\oplus} A(L_{\alpha\beta})$ arises.

\subsection{Courant algebroids and WZNW-Poisson manifolds.}\label{s-WZ-P}
In this section we made a remark that WZNW-Poisson condition
of Klimcik and Strobl \cite{Klim} (formula \ref{wz}below)
 coincides with the condition
that graph of bivector $\pi$ is subalgebra in twisted
Courant algebroid (see Lemma \ref{wz-cour} below).

In the paper \cite{Klim} the authors considered
generalization of Poisson $\sigma$ - model with the help
of the WZNW-type term given by the three form $H$.
They found the consistency condition
which should be imposed on the bivector $\pi$
and closed  three form $H$.
Their  condition is the following:
\bea
[\pi,\pi]_{Schouten}=2<H, \pi \otimes \pi \otimes \pi >
\label{wz}
\eea

(We denote by $ <,>$ the contraction of polyvectors and forms).

Contraction in $<H, \pi \otimes \pi \otimes \pi >$
 is given with respect to the first,  third
and fifth entry.

\subsubsection{Example (due to N. Markarian):}
let us take some nondegenerate, but possibly not
closed 2-form $\omega$ and let $H=d \omega$.
Take $\pi = \omega  ^{(-1)}$
then the pair $\pi, H$ obviously  satisfies WZNW-Poisson condition.

\subsubsection{}

Let us denote by the $\cQ_{H}$ the exact Courant algebroid
obtained from the standard one by twisting
the bracket by the three form $H$,
see subsection \ref{Deform}.

Here we observe the following:

\subsubsection{Lemma} \label{wz-cour} The graph of the mapping
 $ \Omega ^1 \to \cT$ given by the bivector
$\pi$
by the rule $ \alpha \mapsto <\pi, \alpha> $
is {\it subalgebra} in the exact Courant algebroid $\cQ_{H}$
if and only if the WZNW-Poisson equation \ref{wz} is  satisfied.

This lemma is obviously generalization of the original Courant's
theorem which states that Poisson bracket can be characterized
by the property that its graph is subalgebra
in standard Courant algebroid.

The lemma is equivalent to the following
formula:
\bea
&&[< \pi, \alpha >, < \pi, \beta >]=
< [\pi, \pi], \alpha \wedge \beta >
+ < \pi , \cL_{<\pi,\alpha>}\beta>  - \nn \\
&&
- < \pi , \cL_{<\pi,\beta>}\alpha>  -
<\pi , d < \pi, \alpha  \wedge \beta > >
\eea

This formula can be obtained applying two times the
formula \ref{der1} to the LHS ( see example \ref{der-ex}),
then one remarks that RHS can be transformed
by the formula
 $< \pi , \cL_{<\pi,\alpha>}\beta>    -
 d < \pi, \alpha  \wedge \beta >$ $=
< <\pi, \alpha>, d \beta> $
and finally  note that:
$<\pi, \alpha > <\pi, d \beta > = < \pi, < <\pi, \alpha>,d \beta>>$.

\subsubsection{Remark:}
The antisymmetric  bracket in standard Courant algebroid
on the graph of Poisson bivector $\pi$ can be rewritten as follows:
\bea
[< \pi, \alpha >, < \pi, \beta >]_{Cour}=
([ < \pi, \alpha >, < \pi, \beta >]_{vect}, [ \alpha,
\beta]_{Poisson})
\nn
\eea

Where $[,]_{Poisson}$ - is the usual bracket on
1-forms \cite{Pois} defined by the rules $ [df, dg]=d \{ f,g \} $,
and Leibniz rule for functions$ [hdf, dg]=\{ h,g \} df+h[df,dg]$;
$[f, dg]= \{ f,g \} $.
which can be rewritten explicitly as follows:
$[ \alpha, \beta] =  \cL_{<\pi,\alpha>}\beta
-  \cL_{<\pi,\beta>}\alpha  -
 d < \pi, \alpha  \wedge \beta> $.

{\it Remark:} The previous remark can be reformulated
 saying that the map  $\pi: \Omega^1 @>>> \cQ$, given
by $\alpha \mapsto (< \pi, \alpha>, \alpha)$
respects the brackets i.e.  $[,]_{Poisson}$ on  $ \Omega^1$
and antisymmetric Courant's bracket on $\cQ$.
This essentially means that:
$ [ < \pi, \alpha >, < \pi, \beta >]_{vect}=
< \pi ,[ \alpha,\beta]_{Poisson}>
$


\section{Algebroids on the loop spaces and Poisson $\sigma$-model.}
\label{LoopCour}

This section originated from the reading the papers
\cite{Klim, LO} related to Poisson $\sigma$-model
and to some extent the papers \cite{Sch1, Sch2}
We will try to give more geometric and coordinateless reformulation
of some of their formulas.

 We remark that the  phase space of
 Poisson $\sigma$-model is just
cotangent bundle to the loop space of the manifold.
This may be interesting for the geometrically
oriented reader, because the coordinate description
may seems to be quite artificial and this coordinateless
description seems to be very natural.

In the papers \cite{LO} and in some sense in \cite{Klim}
some algebroids played an important role.
These algebroids were constructed from the physical reasons
not completely clear to us, they played role
of symmetries in  Poisson $\sigma$-models.
We propose the construction that starting from
the Lie algebroid on the manifold one can construct
Lie algebroid on  the loop space of the manifold.
(The construction does not work for the Courant algebroids.
Courant algebroid should give Picard-Lie algebroid
on the loop space, which is obvious on the level
of the characteristic classes, but at the moment
we do not know the direct construction).
We remark that after the choice of coordinates,
considering the cotangent bundle of Poisson manifold
as Lie algebroid the constructed algebroid on the loop space
is very similar to the  algebroid considered in \cite{LO}.
The natural generalization of this is to consider
 the Lie algebroid which originates from the
Dirac structures in the Courant algebroid,
we hope that considering the
Dirac structure coming from WZNW-Poisson condition
(see section \ref{s-WZ-P}) and applying to it
our construction one obtains the same Lie algebra
as in \cite{Klim}.

This section organized as follows  first we
explain the construction of Lie algebroid
on the loop space, starting from the one on the manifold itself,
after we give geometric reformulation of Poisson $\sigma$-model
and explain the relation of the first construction to the formulas
from the paper \cite{LO}.

\subsection{From Lie-algebroid on the manifold to
the Lie-algebroid on the loop space.}
We start from the recollection of some well-known
facts about the loop spaces and more generally about the space
of maps from one manifold to another.
We work with smooth manifolds, although
most everything extends to the algebraic situation.

\subsubsection{}
The tangent bundle to the $Map(X,Y)$ equals to
the maps into tangent bundle $Map(X, TY)$.
Note that it is obviously not
true for the cotangent bundle: $Map(X,T^*Y)$ NOT equal
to $T^* Map(X,Y)$.
Also $\Lambda ^n T Map(X,Y)$ not equal to
$ T Map(X, \Lambda ^n Y)$.

\subsubsection{}

There is an obvious map  from the  vector fields on Y
 to the vector fields on $Map(X,Y)$, geometrically
this means that we can move the image of $X$
by the flow generated by vector field $\xi$ on $Y$.
Or  this can be said formally that considering
the composition of $f: X @>>> Y$ and $ \xi: Y @>>> TY$
we obtain the map $X @>>> TY$ and hence the element
of the $T Map(X,Y)$.

{\it Lemma:} this correspondence is obviously
homomorphism of Lie algebras.

\subsubsection{} \label{mult}
There is obviously map $\bbC\times TY \to TY$,
 (just the multiplication of vector fields by scalars)
hence one has map $Map(X, \bbC) \times Map(X,TY)  \to Map ( X,TY)$.
We will denote this multiplication of vector
fields on $Map(X,Y)$ by functions on $X$
as $f(x)*\xi$, in order to distinct it from the
multiplication by functions on $Map(X,Y)$.

{\it Lemma:}
for the vector fields $\xi_i$ on  $ Map(X,Y)$ which
came from the vector fields on $Y$
the following holds:
 $[f_1 (x)* \xi_1, f_2(x)* \xi_2]=
(f_1(x)f_2(x))* [ \xi_1,  \xi_2]$, where
 $f_i$ are  functions on $X$ and the multiplication is described
above.

{\it Remark:}
the lemma above is obviously false without assumption
that  $\xi_i$ came from the vector fields on $Y$.

The lemma is analogous to the following:
$[\sum_{i}a_i f_i (t_i)\frac{\partial}{\partial t_i},
\sum_j b_j h_j (t_j)\frac{\partial}{\partial t_j}]=
\sum_i a_i b_i [  f_i (t_i)\frac{\partial}{\partial t_i},
 h_i (t_i)\frac{\partial}{\partial t_i}]
$
for the vector fields
$\sum_{i} f_i (t_i)\frac{\partial}{\partial t_i}$
and
$\sum_{i} h_i (t_i)\frac{\partial}{\partial t_i}$
where function $f_i$ and $h_i$
depends only on $t_i$.

{\it Remark:} informally this can be expressed as follows:
the vector fields on $Map(X,Y)$ which came
from the vector fields on $Y$
can be written
as $ \int_X F(Y(x)) \frac{ \delta}{\delta Y(x)}$.
In complete analogy with $\sum_i f(Y_i ) \frac{ \partial}{\partial Y_i }$
but the index $i$ is substituted by the continuous index $x$.

{\it Remark:} this multiplication by function on $X$
does not respect the subalgebra of vector fields
which came from $Y$.

{\it Remark:} another way to describe such multiplication
comes from lemma \ref{pr-ev},
one will see that $TMap(X,Y)=pr_*ev^*TM$
and $ev^*TM$ obviously has  structure of $\cO(X)$ module.

\subsubsection{Lemma:}\label{pr-ev}

Consider the diagram:
\bea
\begin{CD}\label{pr-ev-diag}
 Map(X,Y) & \times & X  @>ev>> Y \\
 @V{\pr}VV   @V{\pr_{X}}VV  \\
Map(X,Y) &&  X     \\
\end{CD}
\eea
where $ev$ - is the evaluation map.

Then the following holds true (in the next subsection we will say
the same more explicitly):
\begin{enumerate}
\item
$ev^* TX$ has natural structure of  Lie algebra
(pay attention that here
we used inverse image as a bundle i.e. as sheaf of
$\cO$-modules, this wrong-way functoriality for the vector fields
 is due to
the map $TY \to TMap(X,Y)$).
\item
One can see that $pr_* ev^* TX= TMap(X,Y)$
(this is obviously not true if one puts
any other space instead of $Map(X,Y)$ in the diagram \ref{pr-ev-diag}
this property is characteristic for the space $Map(X,Y)$).

\item
The map $pr^*$ is isomorphism of Lie algebras
$ev^* TX \to  TMap(X,Y) $.

\item
$pr_* (ev ^* T^*Y \otimes pr_X^* {\Omega ^d TX})= T^* Map(X,Y)$,
where $\Omega ^d TX$ - highest forms on $X$
(more exactly one should consider measures on $X$,
but let us do not go into such details).

\end{enumerate}

\subsubsection{ Explicit description of Lie algebra structure
on $ev^*TY$ and its action on $\cO(LM)$.}

By definition $ev^* TY$ is
$\cO( Map(X,Y) \times Y) \underset{\cO(Y)}{ \otimes}ev^{-1} TY$
hence its elements can be written as
$\left(F(\gamma)  \underset{\bbC}{ \otimes} h(x) \right)
\underset{\cO(Y)}{ \otimes}
\xi$, where $\xi$ is vector field on $Y$, and $\gamma \in Map(X,Y)$,
and $F,h$ are functions on $Map(X,Y) $ and $ X$
respectively.

Let us write the formulas corresponding to the lemma above:

\begin{enumerate}
\item

The item 3 in previous lemma stated the isomorphism
 $ev^* TX \to  TMap(X,Y) $.
So we should explain the action of
$ \left(F(\gamma)  \underset{\bbC}{ \otimes} h(x) \right)
\underset{\cO(Y)}{ \otimes}
\xi $ on functions on $Map(X,Y)$.

\bea \label{yavn-act}
 \Bigl(
\left(F(\gamma)  \underset{\bbC}{\otimes} h(x) \right)
\underset{\cO(Y)}{ \otimes}
\xi \Bigr) \Phi(\gamma) =
F(\gamma )
\left( (h(x)*\xi )[  \Phi(\gamma ) ] \right)   \\
\mbox{one multiplies  $(h(x)* \xi)$ as described in subsection
\ref{mult} } \nn
\eea

\item
The multiplication
of vector fields on $Map(X,Y)$ on functions $f(x) \in \cO(X)$
(see section \ref{mult}) in this notations is obviously written as:
\bea
f(x)* ( \left(F (\gamma)  \underset{\bbC}{ \otimes} h(x) \right)
\underset{\cO(Y)}{ \otimes}
\xi )=\left(F (\gamma)  \underset{\bbC}{ \otimes}f(x) h(x) \right)
\underset{\cO(Y)}{ \otimes}
\xi
\eea

\item

The commutator on $ev^* TY$  can be written explicitly as follows
\bea
\label{yavn-com}
[ \left(F_1(\gamma)  \underset{\bbC}{ \otimes} h_1(x) \right)
\underset{\cO(Y)}{ \otimes}
\xi_1  ,
\left(F_2(\gamma)  \underset{\bbC}{ \otimes} h_2(x) \right)
\underset{\cO(Y)}{ \otimes}
\xi_2  ]
= \nn \\
 \left(F_1(\gamma) F_2(\gamma)  \underset{\bbC}{ \otimes}
h_1(x) h_2(x) \right)
\underset{\cO(Y)}{ \otimes}
[\xi_1, \xi_2]  + \nn \\
\left(F_1(\gamma) ( (h_1* \xi_1)[F_2] (\gamma))  \underset{\bbC}{ \otimes} h_2(x) \right)
\underset{\cO(Y)}{ \otimes}
\xi_2  -\nn \\
\left(F_2(\gamma) ( (h_2* \xi_2)[F_1] (\gamma))  \underset{\bbC}{ \otimes} h_1(x) \right)
\underset{\cO(Y)}{ \otimes}
\xi_1
\eea

The expression above is well-defined with respect
to the tensor product taken over $\cO (Y)$.

{\it Remark:}
the formula \ref{yavn-com} is quite obvious the only worth noting
remark is that looking on the LHS
it  may be tempting to write the  wrong expression in the RHS:

$
\left(F_1(\gamma) ( (\xi_1)[F_2] (\gamma))  \underset{\bbC}{ \otimes} h_1 h_2(x) \right)
\underset{\cO(Y)}{ \otimes}
\xi_2
$

The simple form of the first term in LHS is due to the
lemma \ref{mult}.

{\it Remark:}
one can prove the formula
\ref{yavn-com}  as follows:
first
\bea
[ (1 \underset{\bbC}{ \otimes} f_1(z) )
\underset{\cO(Y)}{ \otimes} \xi_1 ,
  (1 \underset{\bbC}{ \otimes} f_2(z) )
\underset{\cO(Y)}{ \otimes} \xi_2 ] =
(1 \underset{\bbC}{ \otimes} f_1(x) f_2(x) )
\underset{\cO(Y)}{ \otimes} [\xi_1 , \xi_2 ]
\eea
this is just the reformulation of the lemma in section \ref{mult}.
The rest part of the formula follows from
the Leibniz rule with respect to the  multiplication
on function $F(\lambda) \in \cO(Map(X,Y))$
and formula \ref{yavn-act}
of the explicit description of the actions
of vector fields on $Map(X,Y)$ on functions on $Map(X,Y)$.

\item The elements of
$ (ev ^* T^*Y \otimes pr_X^* {\Omega ^d TX})$
can be written as

$ (F(\gamma)  \underset{\bbC}{ \otimes}
\omega(x)) \underset{\cO(Y)}{ \otimes}
\alpha(y)$.

The pairing between
$T^* Map(X,Y)=pr_* (ev ^* T^*Y \otimes pr_X^* {\Omega ^d TX})$
and $T Map(X,Y)$ can be given explicitly
as

$ < (F_1(\gamma)  \underset{\bbC}{ \otimes}
\omega(x)) \underset{\cO(Y)}{ \otimes}
\alpha(y) ; (F_2(\gamma)  \underset{\bbC}{ \otimes}
f(x)) \underset{\cO(Y)}{ \otimes}
\xi(y) >=$

$  = F_1(\gamma) F_2(\gamma) \int_X (f(x) <\alpha ;
\xi>_{\gamma(X)} \omega(x)) $.

\item
$T Map(X,Y)$ acts by Lie derivatives on $T^* Map(X,Y)$,
the action can be described explicitly
as follows:

\bea
 \left(F_1(\gamma)  \underset{\bbC}{ \otimes} h(x) \right)
\underset{\cO(Y)}{ \otimes}
\xi_1  \left[ \left(
F_2(\gamma)  \underset{\bbC}{ \otimes} \mu(x) \right)
\underset{\cO(Y)}{ \otimes}
\alpha \right]=\nn \\
F_1 ((h*\xi) F_2)  \underset{\bbC}{ \otimes}\mu(x)
\underset{\cO(Y)}{ \otimes}
\alpha +
F_1 F_2  \underset{\bbC}{ \otimes}h(x)\mu(x)
\underset{\cO(Y)}{ \otimes}
\xi [\alpha]
\eea

\end{enumerate}


\subsubsection{$pr_*ev^*$ of an algebroid is an algebroid.}
Consider the diagram:
\bea
\begin{CD}\label{pr-ev-diag2}
 Map(X,Y) & \times & X  @>ev>> Y \\
 @V{\pr}VV   @V{\pr_{X}}VV  \\
Map(X,Y) &&  X     \\
\end{CD}
\eea
where $ev$ - is the evaluation map.

{\it Lemma:}
Let $A$ be an Lie  algebroid on $Y$,
with the anchor $p : A \to \cT_Y$, then
there is well-defined bracket on $ev^*A$,
which satisfy Jacobi  identity  given by the
formula:
\bea
 [ \left(F_1(\gamma)  \underset{\bbC}{ \otimes} h_1(x) \right)
\underset{\cO(Y)}{ \otimes}
a_1  ,
\left(F_2(\gamma)  \underset{\bbC}{ \otimes} h_2(x) \right)
\underset{\cO(Y)}{ \otimes}
a_2 ]
= \nn \\
 \left(F_1(\gamma) F_2(\gamma)  \underset{\bbC}{ \otimes}
h_1(x) h_2(x) \right)
\underset{\cO(Y)}{ \otimes}
[a_1, a_2]  + \nn \\
\left( F_1(\gamma)  (h_1* p(a_1))[F_2](\gamma)
 \underset{\bbC}{ \otimes} h_2(x) \right)
\underset{\cO(Y)}{ \otimes}
a_2  -\nn \\
\left(F_2(\gamma)  (h_2* p(a_2))[F_1] (\gamma)
\underset{\bbC}{ \otimes} h_1(x) \right)
\underset{\cO(Y)}{ \otimes}
a_1
\eea

One sees that definition is just
copied from the explicit formulas for the
commutator of vector fields on the loop space
(see formula \ref{yavn-com}).
But for general algebroids we just take
this formulas as a definition
and at the moment we do not see geometric
sense of the formulas above.

The main thing to check is to check that the bracket
is well-defined, this can be done directly
using  the property:
assume $\phi(y)$ is function on $Y$
and $\sum_k \phi_1^k \otimes \phi_2^k$ is its pullback
on $Map(X,Y)\times X$,
assume $\xi$ is vector field on $Y$
and we will denote in the same way the corresponding
field in $Map(X,Y)$,
$f(x)$ - any function on $X$
then the following is true:
\bea
\sum_k ((f*\xi)[\phi_1^k]) \otimes \phi_2^k =
\sum_k (\xi[\phi_1^k]) \otimes f\phi_2^k
\eea

It may be tempting to say that it's true
for any $\xi \in \cT_{Map(X,Y)}$ and any function
$\phi \in \cO(Map(X,Y)\times X)$, but it is not the case.

{\it Remark:} note that bracket is not well-defined,
for the $ev^*$ of the Courant algebroid, because
of the absence of the Leibniz rule for the Courant
algebroid. One can see the
problem even in the simplest case:

$  [ (1 \underset{\bbC}{ \otimes} 1 )
\underset{\cO(Y)}{ \otimes}
q_1 ,  (1 \underset{\bbC}{ \otimes} 1 )
\underset{\cO(Y)}{ \otimes}
q_2 ]= (1 \underset{\bbC}{ \otimes} 1 )
\underset{\cO(Y)}{ \otimes} [q_1,q_2] $

This formula leads to the contradictions.

So we have proved the following theorem:
\nopagebreak
\subsubsection{Theorem:}
Let $A$ be an Lie  algebroid  on $Y$,
with the anchor $p : A \to \cT_Y$,
then $pr_*ev^* A $ is an algebroid
on $Map(X,Y)$ with the bracket given above and
the anchor
$\bar p : pr_*ev^* A \to  pr_*ev^* TY$
(recall that $ pr_*ev^* TY= TMap(X,Y)$)
given by the formula:

\bea
  \bar p \left(F(\gamma)  \underset{\bbC}{ \otimes} h(x) \right)
\underset{\cO(Y)}{ \otimes}
a  =
 \left(F(\gamma)  \underset{\bbC}{ \otimes} h(x) \right)
\underset{\cO(Y)}{ \otimes}
p(a)
\eea

\subsubsection{Virasoro algebra and Courant algebroid on the $S^1$
(after Severa)}

{\it Remark:} The natural  quotient of the
standard Courant on
$S^1$ is the Virasoro Lie algebra i.e.
consider
$
0 @>>> \Omega^1 @>>> \cQ @>{\pi}>> \cT @>>> 0
$
And consider the quotient of the  $\Omega^1$ and $ \cQ$
by the space of the exact forms: one obtains
exact sequence:
$
0 @>>> \bbC @>>> Vir  @>{\pi}>> \cT @>>> 0
$
where Vir is Lie algebra (not just Leibniz algebra) because
$[q_1,q_2]+[q_2,q_1]$ - is the exact form.

\subsection{Poisson $\sigma$-model.}

\subsubsection{Lagrangian description of Poisson $\sigma$-model.}
One starts with some manifold M (called target space (TS)) equipped with
 Poisson bivector $\pi$
 and the unit disc $ L = {|z|<1}$ (called worldsheet (WS)).
There are the fields of two kinds in the model:
first maps $ X: L \to M ~~ X(z, \bar z) =(X_1, ... , X_n)$,
second $\xi$ which are  1-forms on $L$ with values
 in the pullback by the map $X$ of the cotangent bundle
to $M$, i.e.  $\xi \in \Gamma( \cT^*_L
\otimes X^* \cT^*_M)$ or explicitly
$\xi(z, \bar z) =(\xi_1, ... , \xi_n)
 ~~~\xi_{k,z}dz + \xi_{k, \bar z}d \bar z$.

The action functional in the model
is given by:
$ S(X, \xi)= \int_L \xi_j d X^j + \pi_{mn} \xi_m \xi_n $.

Or the same can be rephrased in coordinateless
fashion as:
$ S(X, \xi)= \int_L <\xi ,  d^t X > +
< \xi \otimes \xi, X^* (\pi)> $.

Where we mean by $d^t X$ the differential of the map $X$
considered as the section of $ \Gamma( \cT^*_L
\otimes X^* \cT_M)$ so we have natural pairing with
the values in 2-forms on $L$ between $d^t X$ and $\xi$;
in the second term we mean the pairing of
$\xi \otimes \xi $
with  the pullback by the map $X$ of the
bivector $\pi$.

The insight of Kontsevich was that
such model gives  associative star-product.

\subsubsection{Hamiltonian description and
loop space.}

In order to obtain the Hamiltonian description
of the model
one should separate the time variable.
Let (t, $\phi$) be the polar coordinates
and one can take $t$ to be the time variable.
Hence the phase space (the space of fields
when t is fixed)  of the
model  consists of
the maps $X: S^1 \to M$ and $ \xi \in  \Gamma( \cT^*S^1
\otimes X^* \cT_M)$.

{\it Remark:} hence the space of fields
X is just the loop space on M,
it is configuration space, the whole phase space
is cotangent bundle to the loop space
(as it was explained in previous section
$pr_* (ev^* T^*M \otimes T^* S^1) = T^* LM$).

\subsubsection{ Lie algebroid on the loop space and Poisson $\sigma$-model.}




Recall the diagram:
\bea
\begin{CD}
 LM & \times & S^1  @>ev>> M \\
 @V{\pr}VV  \\
   LM       \\
\end{CD}
\eea
where $LM$ is free loop space for M.
In the previous sections we have explained
that there is well-defined Lie algebroid structure
on the $pr_* ev^* A$, where $A$ is any Lie
algebroid on $M$.
Let us take $A=T^*M$ with the Lie algebroid structure
induced by Poisson bivector $\pi$ (see section \ref{Pois-alg}).


{\it Claim:} Lie algebroid  $pr_* ev^* T^*M$
is the same as Lie algebroid of symmetries of Poisson $\sigma$-model
(\cite{LO}
section 5.2 )

Let us explain the claim.
Choose the coordinates $X^i$ on M.
The $1$-forms $dX^i$ gives  trivialization
of the cotangent bundle.
The elements of $pr_* ev^* T^*M$
can be represented as follows:
$\epsilon = \sum_{k=1}^n F_k(\gamma)  \underset{\bbC}{ \otimes} h_k(\phi)
\underset{\bbC}{ \otimes} dX^k$, where $\gamma \in LM$ and $\phi \in S^1$.
Denote by $\epsilon_k(\phi)=F_k(\gamma) h_k(\phi)$. So for the fixed $\phi$ one can consider
$\epsilon_k(\phi)$ as a function on $LM$.
For $\phi \in S^1$ let us denote by $X^k(\phi)$ the function on the loop space
given by $\gamma \mapsto X^k(\gamma(\phi))$ - just the evaluation
of the $k$-th coordinate of the map $\gamma$ at point $\phi$.

{\it Lemma:}
the action of the anchor  $\bar p ( \epsilon)$ on the function
$X^k(\phi)$ is obviously given by the formula:

\bea
\bar p ( \epsilon) (X^k(\phi))= \sum_j \pi^{k,j}(\gamma(\phi))^{k,j} \epsilon_j(\phi)
\eea

This is the same as the formula 5.5 from \cite{LO}
(the anchor was denoted by $\delta_\epsilon$, our map $\gamma$ was denoted
by $X$, the point  $\phi$ is usually omitted in physical notations).
So one sees that our anchor coincides with the anchor for the Lie algebroid
from \cite{LO}.

The  bracket between 1-forms $dX^i \in T^*M$
is obviously given by the formula
$[dX^i,  d X^j]= \partial_{k} \pi^{i,j} dX^k$.
Hence we have the following lemma.

{\it Lemma:}
For the elements $\epsilon \in pr_* ev^* T^*M $ of the special kind:
$\epsilon= \sum_{k=1}^n  1 \underset{\bbC}{ \otimes} h_k(\phi)
\underset{\bbC}{ \otimes} dX^k$
the bracket is obviously given
by the formula:

\bea
[\epsilon, \epsilon' ]=\partial_{k} \pi^{i,j}h_i(\phi)h_j^\prime(\phi)  dX^k
\eea

This is the same bracket which was introduced in \cite{LO}
(see the first formula in section 5.2 of \cite{LO} (the fact that this is the bracket
only on the elements $\epsilon$ of the special kind was somehow not mentioned there)).







In paper \cite{LO} the action of this algebroid on
the space $T^*LM$ was modified by the 1-cocycle (see formula 5.6)
at the moment we do not have an interpretation of this fact
in our approach.

So we have argued that our general and quite simple construction
in particular case gives the same  algebroid as Lie algebroid of symmetries
of Poisson $\sigma$-model,
considered from the other motivation in \cite{LO}.

\subsubsection{Conjectural generalization to WZNW-Poisson $\sigma$-model
and Courant algebroid.}

In analogy to the all above it would be tempting to
propose the following: consider the Courant
obtained by twisting the standard algebroid by the 3-form H
and consider the Dirac structure to corresponding
to WZNW-Poisson condition (see section \ref{s-WZ-P}),
then it is Lie algebroid, one can apply
to it our construction and obtain some Lie algebroid
on the loop space, one can hope that this algebroid
plays the role of symmetries for the WZNW-Poisson $\sigma$-model.

%
\app{}

\sapp{Some formulas.}

The following formulas holds true:
\begin{eqnarray}
[ d,\iota_{\xi_1}\iota_{\xi_2} ]
=\iota_{-[\xi_1,\xi_2]}+\cL_{\xi_1}\iota_{\xi_2}-
 \cL_{\xi_2}\iota_{\xi_1}, \label{K1}\\
{[} d,\iota_{\xi_1}\iota_{\xi_2}\iota_{\xi_3} ]_{+}=
\cL_{\xi_1}\iota_{\xi_2}\iota_{\xi_3}+ c.p.
+\iota_{\xi_1}\iota_{[\xi_2,\xi_3]}+c.p.\label{K2}
\end{eqnarray}

Let us recall that we denoted by $ <,>$
the contraction of polyvectors and forms.
It coincides with $\iota_{\xi} \omega$
for $deg \xi < deg \omega$,
but by conventions
 $\iota_{\xi} \omega = 0$, for $deg \xi > deg \omega$.
And it is not the case for  $ <,>$.
Recall that generalized Lie derivative
$\cL_{\xi}$
action of polyvectors on forms is defined
by the Cartan formula:
$ \cL_{\xi} \omega$ = $ [\iota_{\xi},d]_{graded} \omega$.
Let us modify the definition:

\bea
\cL_{\xi}^{mod}\omega= <\xi,  d \omega> + (-1)^{deg \xi+1} d < \xi,
\omega>
\eea

i.e. we changed the $\iota_{\xi}$ to contraction with $\xi $ :
$<\xi, \bullet >$.

Let us denote by $ [polyvector , polyvector ]$ the usual Schouten bracket
for the polyvectors and
$ [polyvector, form] $ - the  modified Lie derivative
action of polyvectors on forms.

The contraction and commutator are consistent
in the sense that the following holds:

\bea
[ \xi, < \nu, \omega > ] =
< [\xi,  \nu ], \omega > +
< \nu , [\xi, \omega] > \label{der1}
\eea

It is the analog of the usual
$[\cL_{\xi} , \iota_{\nu}]=\iota_{[\xi, \nu]}$.

For example:
\bea
&&
[< \pi, \alpha >, < \pi, \beta >]=
< [\pi, < \pi, \beta>], \alpha> - <\pi, \cL_{< \pi,\beta >} \alpha>
\nn \\
&&
[\pi, < \pi, \beta>]=
< [\pi, \pi ], \beta > + <\pi, [\pi, \beta] > \label{der-ex}
\eea


\end{document}